\documentclass[sigconf]{acmart}
\AtBeginDocument{%
  }
\usepackage{multirow}
\usepackage{booktabs}
\usepackage{array}
\usepackage{amsmath}
\setcopyright{acmlicensed}
\copyrightyear{2025}
\acmYear{2025}
\acmDOI{XXXXXXX.XXXXXXX}
\acmConference[Conference acronym 'XX]{Make sure to enter the correct
  conference title from your rights confirmation email}{June 03--05,
  2018}{Woodstock, NY}
\acmISBN{978-1-4503-XXXX-X/2018/06}

\acmSubmissionID{99}

\begin{document}

\title{Personality-Enhanced Multimodal Depression Detection in the Elderly}

\author{Honghong Wang}
\email{wanghonghong@fosafer.com}
\orcid{0009-0005-3715-7962}
\affiliation{%
  \institution{Beijing Fosafer Information Technology Co., Ltd.}
  \state{Beijing}
  \country{China}
}

\author{Jing Deng}
\email{dengjing@fosafer.com}
\affiliation{%
  \institution{Beijing Fosafer Information Technology Co., Ltd}
  \state{Beijing}
  \country{China}
}

\author{Rong Zheng}
\email{zhengrong@fosafer.com}
\affiliation{%
  \institution{Beijing Fosafer Information Technology Co., Ltd}
  \state{Beijing}
  \country{China}
}


\begin{abstract}
  This paper presents our solution to the Multimodal Personality-aware Depression Detection (MPDD) challenge at ACM MM 2025. We propose a multimodal depression detection model in the Elderly that incorporates personality characteristics. We introduce a multi-feature fusion approach based on a co-attention mechanism to effectively integrate LLDs, MFCCs, and Wav2Vec features in the audio modality. For the video modality, we combine representations extracted from OpenFace, ResNet, and DenseNet to construct a comprehensive visual feature set. Recognizing the critical role of personality in depression detection, we design an interaction module that captures the relationships between personality traits and multimodal features. Experimental results from the MPDD Elderly Depression Detection track demonstrate that our method significantly enhances performance, providing valuable insights for future research in multimodal depression detection among elderly populations.
\end{abstract}

\ccsdesc[500]{Computing methodologies~Artificial intelligence}
\ccsdesc[500]{Applied computing~Health informatics}
\ccsdesc[500]{Information systems~Multimedia information systems}

\keywords{Multimodal Depression Detection, Transformer Fusion, Multilevel Feature}

\maketitle

\section{Introduction}
Depression is a widespread mental health disorder that significantly affects the well-being and quality of life of millions of people around the world \cite{b1}. 
According to the World Health Organization (WHO) , approximately 280 million people worldwide suffer from depression. Data from The Lancet Global Burden of Disease (GBD) study show that between 2010 and 2021, depression became the second leading cause of disability-adjusted life years (DALY)lost, with its global burden continuing to increase \cite{b2}.
Traditional depression screening methods are commonly based on the Patient Health Questionnaire-9 (PHQ-9), which evaluates depressive symptoms using a score ranging from 0 to 27. Scores above 4 suggest the presence of depression, with 5–9 indicating mild, 10–14 moderate, 15–19 moderately severe, and 20–27 severe depression \cite{b3}. 
Traditional depression screening methods are dependent on self-reported responses, making them vulnerable to bias and limiting their effectiveness in capturing the complex and evolving nature of depression, particularly in detecting early or subtle changes in mental state. 

\begin{figure*}[t]  
  \centering
  \includegraphics[width=\textwidth]{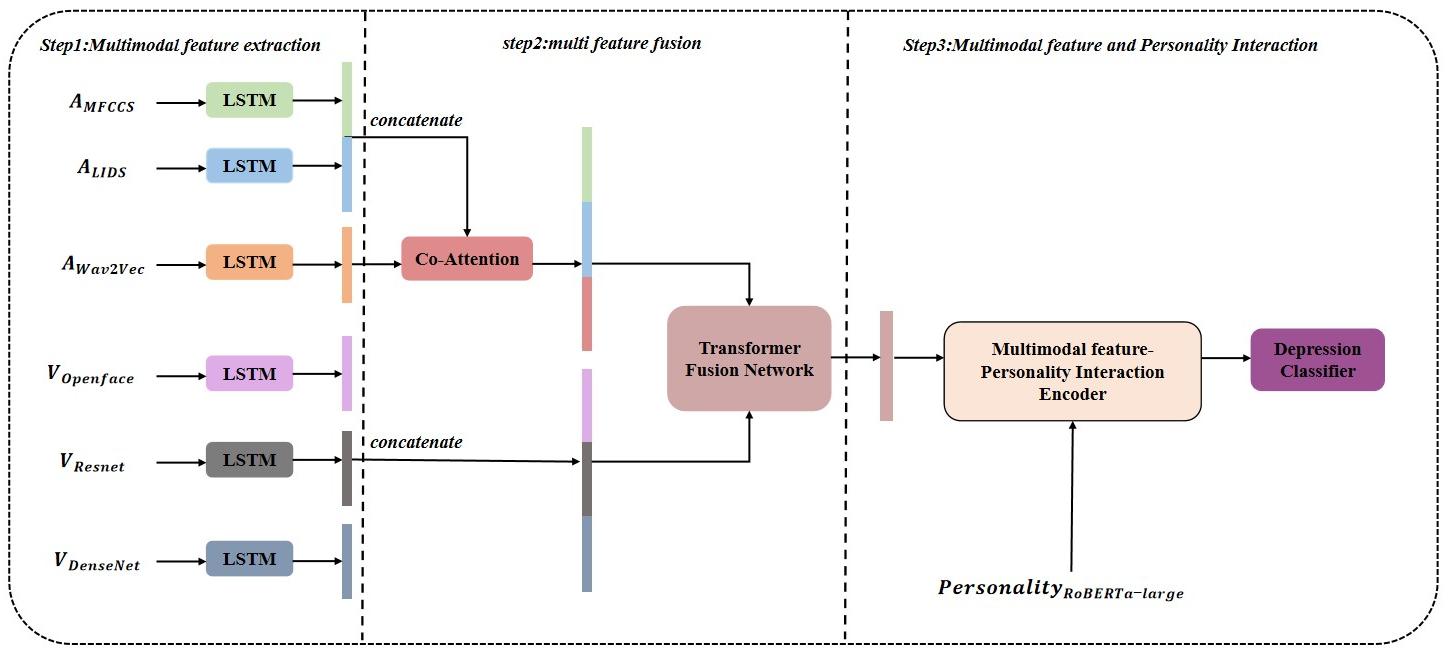} 
  \caption{Overview of our multimodal depression detection framework integrating personality traits.}
  \Description{Overview of our multimodal depression detection framework.}
\end{figure*}

Deep learning has significantly improved the performance of depression detection. Deep learning-based depression detection technologies leverage physiological and behavioral data, including voice, text, video, and Electroencephalogram (EEG) signals, collected from test subjects. These data provide valuable indicators of depression.  Depressed individuals often demonstrate reduced fundamental vocal frequency, exhibit facial expressions associated with sadness and fear, and exhibit thoughts and behaviors consistent with the PHQ-9 items. Consequently, researchers have focused on the detection of depression in various modalities, such as images 
\cite{b4,b5,b6},speech \cite{b7,b8,b9},
and text \cite{b10,b11,b12}.
Although each modality offers valuable information, relying on a single source limits the depth of information captured. Compared to unimodal approaches, multimodal data provide more comprehensive and complementary signals, substantially improving detection performance \cite{b13,b14,b15}.

However, current depression detection algorithms face several critical challenges. First,  most existing studies and publicly available datasets focus primarily on adolescents, with relatively little attention given to the elderly population. This is a critical gap, as depressive symptoms often manifest differently in age groups. For example, studies indicate that older adults with depression typically exhibit lower levels of anger and guilt but higher levels of anxiety compared to younger individuals \cite{b16,b17}.

Second, in the audio modality, recent studies have primarily relied on single feature types independently, such as low-level descriptors (LLDs) extracted by OpenSMILE \footnote{https://github.com/audeering/opensmile}, handcrafted features such as mel-spectrograms, or high-level embeddings from pretrained models. This limitation also applies to other modalities, such as video and text. However, using a single type of feature often fails to capture the full complexity of audio signals, limiting the model's ability to detect depression-related cues effectively. This highlights the need for richer and more integrated multi-level representations in multimodal depression detection.

Third, existing depression detection research often overlooks the influence of personality traits. Studies have identified strong correlations between certain personality traits and the onset of depression \cite{b18,b19}.
Steunenberg et al. 
\cite{b20} conducted a 6-year longitudinal study involving 1,511 elderly individuals without depression at baseline. The study identified three personality traits significantly associated with the later development of depression: increased neuroticism, diminished coping ability, and lower self-efficacy. Individuals exhibiting these traits were found to be 3.6, 2.0, and 1.7 times more likely, respectively, to develop depression compared to the general population.

To address the challenges outlined above, this study proposes a multimodal depression detection framework tailored for the elderly. The method incorporates personality traits to enable adaptive modeling of individual depression characteristics, enhancing personalization and model relevance. Additionally, it integrates multi-level features from both audio and video modalities to capture a more comprehensive representation of depressive behavior. The primary contributions of this study are as follows.

(1) We propose a multimodal depression detection network specifically designed for the elderly. In the audio modality, the model captures multi-level representations from both the time and frequency domains to enrich the characterization of depressive speech. A co-attention mechanism is introduced to effectively fuse these features. For the video modality, we enhance facial representation by concatenating features extracted from diverse tasks and pre-trained models.

(2) We design a Personality Traits and Multimodal Feature Interaction Module (PTMFIM) to model the deep correlation between depressive characteristics and individual personality traits. This enables the network to adapt more effectively to personalized patterns of depression expression.

\begin{figure*}[t]
  \centering
  \includegraphics[width=\linewidth]{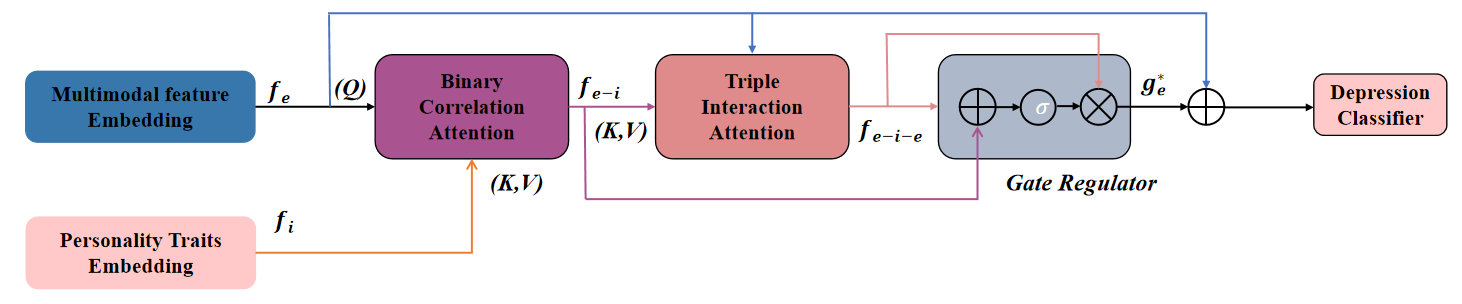}
  \caption{Multimodal feature and personality traits iteration module.}
  \Description{A woman and a girl in white dresses sit in an open car.}
\end{figure*}

\section{Proposed Model}
The proposed framework is illustrated in Figure 1. The model consists of two branches, with the audio branch extracting three levels of acoustic features to capture complementary aspects of speech. These features are encoded and integrated using a co-attention mechanism to enable multi-granularity fusion. In the video branch, an encoder processes and concatenates visual features obtained at different levels. The resulting audio and visual features are then aggregated into utterance level representations using an Attentive Static Pooling (ASP) module. To fusion cross-modal features, a Transformer Fusion module is employed to fuse the multimodal features. To capture the complex relationships between personality traits and multimodal features, we propose the Personality traits and Multimodal Feature Interaction Module (PTMFIM). PTMFIM is designed to learn the deep interactions between personality embeddings and multimodal representations, enabling more personalized and context-aware depression detection. The PTMFIM output features are then passed to the depression classifier, which predicts the level of depression based on the enriched representation.

\subsection{Multi-level Feature Fusion of Audio and Visual}

In the audio branch, we first used the open-source tool OpenSMILE to extract LLDs from raw speech, capturing basic acoustic properties such as short-term energy and zero-crossing rate. In addition, Mel-Frequency Cepstral Coefficients (MFCCs) \footnote{https://github.com/librosa/librosa}, widely adopted in speech processing, were extracted to represent the short-term spectral characteristics of audio signals. To further enrich the representation, we utilized the pre-trained Wav2Vec model \footnote{https://github.com/facebookresearch/fairseq/tree/main/examples/wav2vec} to extract high-dimensional audio embeddings. Trained on a large corpus of unlabeled audio, Wav2Vec provides informative representations that encapsulate complex acoustic patterns. All three types of features—LLDs, MFCCs, and Wav2Vec embeddings—were encoded using an LSTM encoder to preserve temporal dependencies and prepare them for subsequent fusion. Building on the diverse auditory features that are known to contribute uniquely to depression detection and guided by insights from 
\cite{b21}, we propose a Co-Attention module to effectively integrate Wav2Vec, LLDs and MFCC features. The integration process begins with non-linear transformations of each feature type using linear layers combined with Dropout and ReLU activations. Next, the LLDs and MFCCs features are concatenated along the feature dimension. Since Wav2Vec provides a high-level speech representation, these concatenated features are used to weight the Wav2Vec features through element-wise multiplication. Finally, the weighted Wav2Vec features are concatenated with the LLD and MFCC features to form a comprehensive audio representation.

For the video modality, facial features are first extracted from raw video frames using the OpenFace toolkit \footnote{http://multicomp.cs.cmu.edu/resources/openface/}. To further enrich the representation, we also extract complementary features using pre-trained DenseNet \footnote{https://github.com/liuzhuang13/DenseNet} and ResNet \footnote{https://huggingface.co/microsoft/resnet-50}architectures, both widely used in image classification tasks. Finally, the outputs from all three sources are concatenated to form a comprehensive visual representation.

\subsection{Personality and multimodal feature interaction module}

For the multimodal representations extracted in Section 2.1, we initially employed ASP to transform frame level features into utterance level features. Subsequently, the Transformer Fusion module was employed to fuse visual and audio features. Assuming that $u_{n}^{v}$ and $u_{n}^{a}$ represent visual and acoustic features of the \textit{n}-th sample, the multimodal representation $f_{s}^{*}$ can be expressed as follows:

$$
f_{s}^{*} = F_{s}^{*} \left( \operatorname{Concat} \left( F_{s}^{v} \left( u_{n}^{v} \right), F_{s}^{a} \left( u_{n}^{a} \right) \right) \right)
$$

Where $F_{s}^{a}$ is the Acoustic Encoder based on LSTM and ASP, $F_{s}^{v}$ is the Visual Encoder that has the same structure as $F_{s}^{a}$. $F_{s}^{*}$ represent the Transformer fusion module, composed of multiple layers of Transformer encoder. The concatenated audio and visual sequence features are fed into the network, where self-attention is applied to capture long-range dependencies across modalities. The output is a unified depression representation enriched with global contextual information.

To extract human personality traits, the initial step involves the construction of a patient prompt. This prompt is derived from the patient's Big5 scores, age, gender, place of origin, and other relevant demographic information. The specific prompt is as follows:

\textit{The patient is a 50 male from Beijing. 
The patient's Extraversion score is extroversion. 
The Agreeableness score is agreeableness. 
The Openness score is openness. 
The Neuroticism score is neuroticism. 
The Conscientiousness score is conscientiousness. 
Please generate a concise, fluent English description summarizing the patient's key personality traits, family environment, and other notable characteristics. 
Avoid mentioning depression or related terminology. 
Output the response as a single paragraph.}

Subsequently, the ChatGLM3 Chinese large language model\footnote{\url{https://github.com/THUDM/ChatGLM3}} is employed to generate a personalized textual description based on the constructed prompt. This description is then encoded using the RoBERTa-large model to obtain a personalized depression text embedding.

Inspired by \cite{b21},
we propose the PTMFIM, which models the complex correlations between personality traits and multimodal features to learn deep personalized interaction patterns. As shown in Figure 2, the proposed module consists of three key components: Binary Correlation Attention, Triple Interaction Attention, and a Gating Regulator. The Binary Correlation Attention component uses a cross-attention mechanism to capture pairwise relationships between personality traits and multimodal representations. To enable this, a linear projection layer first maps both personality traits and multimodal features to the corresponding query (Q), key (K) and value (V) vectors. The attention mechanism is then applied to capture the underlying correlations between personality traits and multimodal features. Building on this, the Triple Interaction Attention component further models higher-order interactions among personality traits and fused audio-visual features, enabling a deeper understanding of their joint influence on depression detection. The Cascade Interaction Representation is computed by using personality traits as query vectors, while the output from the Binary Correlation Attention serves as the key and value vectors. Finally, the Gate Regulator automates the learning of both binary and higher-order (ternary) interaction features by introducing a gating mechanism that models the influence of relationships between personality traits and multimodal representations. This component allows the model to dynamically adjust the contribution of personality and multimodal features to depression detection. Specifically, a sigmoid function is used to compute a control gate value based on the output of the binary correlation attention and triple interaction attention modules. This gate value is then applied to the output of the Triple Interaction Attention through element-wise multiplication. The gated output is subsequently combined with the personality feature embedding to produce the final output of the interaction module.

\section{EXPERIMENTS and RESULTS}

\subsection{Datasets}

We conduct a series of experiments using data from the Multimodal Personality-Aware Depression Detection (MPDD) challenge \footnote{https://hacilab.github.io/MPDDChallenge.github.io/} held at the 2025 ACM Multimedia Conference. The dataset, processed with 1s and 5s sampling windows for feature extraction, yields 564 samples—337 for training and validation, and 227 for testing. Neither 1s nor 5s indicates the length of the sample. instead, they refer to the size of the hopping window used during the preprocessing stage. Due to the limited amount of training data, a resampling strategy is employed during training. To address sample variability, we randomly partition the training and validation sets to ensure a representative distribution of the data. The MPDD defines three classification tasks based on depression severity. These tasks correspond to binary, ternary, or quinary classification schemes, depending on the criteria used. The binary classification system distinguishes between healthy and depressed states. The ternary classification system involves the additional delineation of normal, mild, and severe depression. The quinary classification system, on the other hand, incorporates an additional level of complexity by distinguishing between normal, mild depression, moderate depression, moderate severity, and severe depression. Performance is evaluated by averaging weighted and unweighted precision, as well as weighted and unweighted F1 scores, offering a comprehensive measure of the effectiveness of the model in multiclass depression detection. The final metric for each task is the average of weighted and unweighted values:
\[
\begin{aligned}
\textit{Acc}^{\textit{task}} &= \frac{\textit{Acc}_{\textit{weighted}} + \textit{Acc}_{\textit{unweighted}}}{2} \\
F_{1}^{\textit{task}} &= \frac{F_{1,\textit{weighted}} + F_{1,\textit{unweighted}}}{2}
\end{aligned}
\]

\begin{table}
  \caption{Results of the proposed method compared to the baseline model, where baseline is above and proposed is below }
  \label{tab:freq}
  \begin{tabular}{cccc}
    \toprule
    Length & Task & Accuracy & F1-score \\
    \midrule
    \multirow{6}{*}{1s} 
        & Binary   & 85.01 & 82.42 \\
        &          & 89.18 & 88.53 \\
    \cmidrule(lr){2-4}
        & Ternary  & 55.95 & 56.06 \\
        &          & 59.16 & 58.53 \\
    \cmidrule(lr){2-4}
        & Quinary  & 57.51 & 47.53 \\
        &          & 61.33 & 60.67 \\
    \bottomrule
    \multirow{6}{*}{5s} 
        & Binary   & 78.09 & 77.06 \\
        &          & 79.81 & 78.67 \\
    \cmidrule(lr){2-4}
        & Ternary  & 58.66 & 58.80 \\
        &          & 63.99 & 62.62 \\
    \cmidrule(lr){2-4}
        & Quinary  & 67.96 & 67.01 \\
        &          & 72.17 & 72.48 \\
    \bottomrule
  \end{tabular}
\end{table}

Table 1 presents a comparative analysis of our proposed approach against the baseline model provided by MPDD on the test set. The baseline employs a single feature from each audio and video modality, concatenate personality traits features directly with multimodal representations. The best-performing results from the audio and video feature permutation experiments are selected as the baseline result. Our method demonstrates significant improvements over the baseline across all three multi-class classification tasks, for both 1s and 5s sampling window lengths. In the three classification tasks using 1s sampling window, our model outperformed the baseline with average accuracy improvements of 4.17\%, 3.21\%, and 3.82\%, and F1-score gains of 6.11\%, 2.47\%, and 13.14\%, respectively. With 5s sampling window, the model achieved accuracy improvements of 4.41\%, 5.9\%, and 4.21\%, along with F1-score increases of 7.9\%, 13.8\%, and 0.47\%. These results demonstrate the effectiveness of our proposed method.

\subsection{Ablation Study on Highlighted Modules}

\begin{table}
\centering
\caption{Results of ablation experiments with 1s sampling window}
\label{tab:results}
\begin{tabular}{>{\raggedright\arraybackslash}p{2.5cm}c>{\centering\arraybackslash}p{1.5cm}>{\centering\arraybackslash}p{1.5cm}}
\toprule
Method & Task & Accuracy & F1-score \\
\midrule
\multirow{3}{*}{Baseline} 
    & Binary   & 85.01 & 82.42 \\
    & Ternary  & 55.95 & 56.06 \\
    & Quinary  & 57.51 & 47.53 \\
\midrule[0.5pt]

\multirow{3}{*}{w/o Multi-Audio}
    & Binary   & 87.90 & 86.77 \\
    & Ternary  & 59.08 & 58.22 \\
    & Quinary  & 60.00 & 59.77 \\
\midrule[0.5pt]

\multirow{3}{*}{w/o co-att}
    & Binary   & 88.89 & 87.74 \\
    & Ternary  & 58.98 & 58.33 \\
    & Quinary  & 57.73 & 54.24 \\
\midrule[0.5pt]

\multirow{3}{*}{w/o Multi-Visual}
    & Binary   & 88.63 & 87.01 \\
    & Ternary  & 58.88 & 56.00 \\
    & Quinary  & 56.09 & 56.88 \\
\midrule[0.5pt]

\multirow{3}{*}{w/o PTMFIM}
    & Binary   & 86.29 & 85.93 \\
    & Ternary  & 56.93 & 55.35 \\
    & Quinary  & 58.06 & 54.98 \\
\midrule[0.5pt]

\multirow{3}{*}{\textbf{Proposed}}
    & Binary   & \textbf{89.18} & \textbf{88.53} \\
    & Ternary  & \textbf{59.16} & \textbf{58.53} \\
    & Quinary  & \textbf{61.33} & \textbf{60.67} \\
\bottomrule
\end{tabular}
\end{table}

This section validates each proposed component through ablation experiments. Table 2 compares four variants:(1)Without audio multi-feature fusion.
(2)Without Co-Attention in audio feature fusion.
(3)Without visual multi-feature concatenation.
(4)Without PTMFIM.
Table 2 presents the results of the ablation experiments using a 1s sampling window. The quantitative analysis shows that all proposed modules contribute positively to performance. Among the proposed modules, PTMFIM has the greatest impact, increasing average accuracy by 2.89\%, 2.23\%, and 3.27\%, and average F1 scores by 2.60\%, 3.18\%, and 5.69\% across the three classification tasks, respectively. These results underscore the effectiveness of integrating personality features with multimodal representations to enhance depression detection. Additionally, employing multiple audio and video features, along with the co-attention mechanism for audio feature fusion, yielded consistent performance gains over single-feature baselines.

\begin{table}
\centering
\caption{Results of ablation experiments with 5s sampling window}
\label{tab:results_extension}
\begin{tabular}{>{\raggedright\arraybackslash}p{2.5cm}c>{\centering\arraybackslash}p{1.5cm}>{\centering\arraybackslash}p{1.5cm}}
\toprule
Method & Task & Accuracy & F1-score \\
\midrule
\multirow{3}{*}{Baseline} 
    & Binary   & 78.09 & 77.06 \\
    & Ternary  & 58.66 & 58.80 \\
    & Quinary  & 67.96 & 67.01 \\
\cmidrule(lr){1-4} 

\multirow{3}{*}{w/o Multi-Audio}
    & Binary   & 79.30 & 78.34 \\
    & Ternary  & 61.38 & 59.85 \\
    & Quinary  & 70.14 & 69.51 \\
\cmidrule(lr){1-4}

\multirow{3}{*}{w/o co-att}
    & Binary   & 77.30 & 77.53 \\
    & Ternary  & 60.35 & 59.52 \\
    & Quinary  & 68.74 & 68.36 \\
\cmidrule(lr){1-4}

\multirow{3}{*}{w/o Multi-Visual}
    & Binary   & 78.78 & 77.75 \\
    & Ternary  & 59.90 & 59.91 \\
    & Quinary  & 70.25 & 69.29 \\
\cmidrule(lr){1-4}

\multirow{3}{*}{w/o PTMFIM}
    & Binary   & 78.66 & 77.91 \\
    & Ternary  & 59.08 & 58.78 \\
    & Quinary  & 68.30 & 66.82 \\
\cmidrule(lr){1-4}

\multirow{3}{*}{\textbf{Proposed}}
    & Binary   & \textbf{79.81} & \textbf{78.67} \\
    & Ternary  & \textbf{63.99} & \textbf{62.62} \\
    & Quinary  & \textbf{72.17} & \textbf{72.48} \\
\bottomrule
\end{tabular}
\end{table}
 
Furthermore, ablation experiments using a 5s sampling window, as shown in Table 3, confirm that each proposed component contributes to performance gains to varying degrees. Similar to the 1s sampling window, the Co-Attention mechanism for audio feature fusion and the PTMFIM module played a significant role in improving depression detection performance. Furthermore, the Multi-Audio/Visual approach improved the final results, indicating that multi-scale features provide complementary information that is beneficial for depression detection. These results further demonstrate the robustness and effectiveness of our method across different sampling scales.

\section{Conclusion}

This study presents a multimodal depression detection model for the elderly that integrates personality traits. For the audio modality, we introduce a multi-feature fusion module based on a co-attention mechanism that combines three features. For the video modality, three distinct features are concatenated. Additionally, we propose an encoder to model interactions between personality traits and multimodal features. Evaluation on the Multimodal Personality-aware Depression Detection (MPDD) challenge demonstrates that our approach significantly enhances depression detection performance in the elderly population.

\section{Acknowledgments}

This work is supported by the National Key Research and Development Program of China (No.2022YFF0608504).



\bibliographystyle{ACM-Reference-Format}

\begin{thebibliography}{00}
\bibitem{b1} Abdoli N, Salari N, Darvishi N, Jafarpour S, Solaymani M, Mohammadi M, Shohaimi S. The global prevalence of major depressive disorder (MDD) among the elderly: A systematic review and meta-analysis. Neurosci Biobehav Rev. 2022 Jan;132:1067-1073. doi: 10.1016/j.neubiorev.2021.10.041. Epub 2021 Nov 4. PMID: 34742925.
\bibitem{b2} GBD 2021 Demographics Collaborators. Global age-sex-specific mortality, life expectancy, and population estimates in 204 countries and territories and 811 subnational locations, 1950-2021, and the impact of the COVID-19 pandemic: a comprehensive demographic analysis for the Global Burden of Disease Study 2021. Lancet. 2024 May 18;403(10440):1989-2056. doi: 10.1016/S0140-6736(24)00476-8. Epub 2024 Mar 11. PMID: 38484753; PMCID: PMC11126395.
\bibitem{b3} Li T, Wei J, Fritzsche K, Toussaint AC, Zhang L, Zhang Y, Chen H, Wu H, Ma X, Li W, Ren J, Lu W, Leonhart R. Validation of the Chinese version of the Somatic Symptom Scale-8 in patients from tertiary hospitals in China. Front Psychiatry. 2022 Sep 28;13:940206. doi: 10.3389/fpsyt.2022.940206. PMID: 36276338; PMCID: PMC9583900.
\bibitem{b4} He L, Guo C, Tiwari P, et al. Intelligent system for depression scale estimation with facial expressions and case study in industrial intelligence[J]. International Journal of Intelligent Systems, 2022, 37(12): 10140-10156.
\bibitem{b5} Bin Hu, Yongfeng Tao, and Minqiang Yang. 2023. Detecting depression based on facial cues elicited by emotional stimuli in video. Comput. Biol. Med. 165, C (Oct 2023). https://doi.org/10.1016/j.compbiomed.2023.107457.
\bibitem{b6} C. Á. Casado, M. L. Cañellas and M. B. López, "Depression Recognition Using Remote Photoplethysmography From Facial Videos," in IEEE Transactions on Affective Computing, vol. 14, no. 4, pp. 3305-3316, 1 Oct.-Dec. 2023, doi: 10.1109/TAFFC.2023.3238641.
keywords: {Feature extraction;Depression;Heart rate variability;Videos;Faces;Biomedical monitoring;Skin;Affective computing;depression detection;HRV features;image processing;machine learning;rPPG;remote photoplethysmography;signal processing},
\bibitem{b7} Sara Sardari, Bahareh Nakisa, Mohammed Naim Rastgoo, and Peter Eklund. 2022. Audio based depression detection using Convolutional Autoencoder. Expert Syst. Appl. 189, C (Mar 2022). https://doi.org/10.1016/j.eswa.2021.116076.
\bibitem{b8} Kaur B, Rathi S, Agrawal RK. Enhanced depression detection from speech using Quantum Whale Optimization Algorithm for feature selection. Comput Biol Med. 2022 Nov;150:106122. doi: 10.1016/j.compbiomed.2022.106122. Epub 2022 Sep 23. PMID: 36182759.
\bibitem{b9} Y. Li, C. Sun and Y. Dong, "A Novel Audio-Visual Multimodal Semi-Supervised Model Based on Graph Neural Networks for Depression Detection," ICASSP 2025 - 2025 IEEE International Conference on Acoustics, Speech and Signal Processing (ICASSP), Hyderabad, India, 2025, pp. 1-5, doi: 10.1109/ICASSP49660.2025.10888673. keywords: {Representation learning;Visualization;Signal processing;Semisupervised learning;Depression;Feature extraction;Graph neural networks;Data models;Speech processing;Unsupervised learning;audio-visual multimodal;semi-supervised;depression detection;graph neural networks},
\bibitem{b9} Y. Shen, H. Yang and L. Lin, "Automatic Depression Detection: an Emotional Audio-Textual Corpus and A Gru/Bilstm-Based Model," ICASSP 2022 - 2022 IEEE International Conference on Acoustics, Speech and Signal Processing (ICASSP), Singapore, Singapore, 2022, pp. 6247-6251, doi: 10.1109/ICASSP43922.2022.9746569. keywords: {Computer science;Codes;Conferences;Mental health;Linguistics;Signal processing;Depression;Depression detection;Multi-modal fusion;EATD-Corpus},
\bibitem{b10} Sergio G. Burdisso, Marcelo Errecalde, and Manuel Montes-y-Gómez. 2019. A text classification framework for simple and effective early depression detection over social media streams. Expert Syst. Appl. 133, C (Nov 2019), 182–197. https://doi.org/10.1016/j.eswa.2019.05.023.
\bibitem{b11} C. Fu et al., "HAG: Hierarchical Attention with Graph Network for Dialogue Act Classification in Conversation," ICASSP 2023 - 2023 IEEE International Conference on Acoustics, Speech and Signal Processing (ICASSP), Rhodes Island, Greece, 2023, pp. 1-5, doi: 10.1109/ICASSP49357.2023.10096805. keywords: {Semantics;Oral communication;Logic gates;Signal processing;Feature extraction;Graph neural networks;Encoding},
\bibitem{b12} Seneviratne N, Espy-Wilson C. Multimodal depression classification using articulatory coordination features and hierarchical attention based text embeddings[C]//ICASSP 2022-2022 IEEE International Conference on Acoustics, Speech and Signal Processing (ICASSP). IEEE, 2022: 6252-6256.
\bibitem{b13} Anupama Ray, Siddharth Kumar, Rutvik Reddy, Prerana Mukherjee, and Ritu Garg. 2019. Multi-level Attention Network using Text, Audio and Video for Depression Prediction. In Proceedings of the 9th International on Audio/Visual Emotion Challenge and Workshop (AVEC '19). Association for Computing Machinery, New York, NY, USA, 81–88. https://doi.org/10.1145/3347320.3357697.
\bibitem{b14} Shen Y, Yang H, Lin L. Automatic depression detection: An emotional audio-textual corpus and a gru/bilstm-based model[C]//ICASSP 2022-2022 IEEE International Conference on Acoustics, Speech and Signal Processing (ICASSP). IEEE, 2022: 6247-6251.
\bibitem{b15} Hegeman JM, Kok RM, van der Mast RC, Giltay EJ. Phenomenology of depression in older compared with younger adults: meta-analysis. Br J Psychiatry. 2012 Apr;200(4):275-81. doi: 10.1192/bjp.bp.111.095950. PMID: 22474233.
\bibitem{b16} Fiske A, Wetherell JL, Gatz M. Depression in older adults. Annu Rev Clin Psychol. 2009;5:363-89. doi: 10.1146/annurev.clinpsy.032408.153621. PMID: 19327033; PMCID: PMC2852580.
\bibitem{b17} Koorevaar AM, Comijs HC, Dhondt AD, van Marwijk HW, van der Mast RC, Naarding P, Oude Voshaar RC, Stek ML. Big Five personality and depression diagnosis, severity and age of onset in older adults. J Affect Disord. 2013 Oct;151(1):178-85. doi: 10.1016/j.jad.2013.05.075. Epub 2013 Jun 29. PMID: 23820093.
\bibitem{b18} Sadeq NA, Molinari V. Personality and its Relationship to Depression and Cognition in Older Adults: Implications for Practice. Clin Gerontol. 2018 Oct-Dec;41(5):385-398. doi: 10.1080/07317115.2017.1407981. Epub 2017 Dec 26. PMID: 29279022.
\bibitem{b19} Steunenberg B, Beekman AT, Deeg DJ, Kerkhof AJ. Personality and the onset of depression in late life. J Affect Disord. 2006 Jun;92(2-3):243-51. doi: 10.1016/j.jad.2006.02.003. Epub 2006 Mar 20. PMID: 16545466.
\bibitem{b20} H. Zou, Y. Si, C. Chen, D. Rajan and E. S. Chng, "Speech Emotion Recognition with Co-Attention Based Multi-Level Acoustic Information," ICASSP 2022 - 2022 IEEE International Conference on Acoustics, Speech and Signal Processing (ICASSP), Singapore, Singapore, 2022, pp. 7367-7371, doi: 10.1109/ICASSP43922.2022.9747095. keywords: {Emotion recognition;Speech recognition;Feature extraction;Acoustics;Data mining;Speech processing;Task analysis;Speech emotion recognition;Multimodal fusion;Multi-level acoustic information;Co-attention mechanism},
\bibitem{b21} Liu R, Zuo H, Lian Z, et al. Emotion and intent joint understanding in multimodal conversation: A benchmarking dataset[J]. arXiv preprint arXiv:2407.02751, 2024.

\end{thebibliography}

\appendix

\end{document}